\documentclass[a4paper,10pt]{article}
\usepackage{authblk}

\usepackage{graphicx}  

\title{Photoluminescence of nitrogen-vacancy and silicon-vacancy color centers in phosphorus-doped  diamond at room and higher temperatures}

\author[1]{F.~Sledz\thanks{Contributed equally to this work}}
\author[2]{S.~Piccolomo\thanks{Contributed equally to this work}}
\author[1]{A. M.~Flatae\thanks{Contributed equally to this work}}
\author[1,3]{S.~Lagomarsino}
\author[4]{R.~Rechenberg}
\author[4]{M. F.~Becker}
\author[3,5]{S.~Sciortino}
\author[3]{N.~Gelli}
\author[6]{I. A.~Khramtsov}
\author[6]{D. Yu.~Fedyanin}
\author[2]{G.~Speranza}
\author[3,5]{L.~Giuntini}
\author[1,7]{M.~Agio\thanks{mario.agio@uni-siegen.de}}

\affil[1]{Laboratory of Nano-Optics and C$\mu$, University of Siegen  - Siegen, Germany}
\affil[2]{Centro Materiali e Microsistemi, Fondazione Bruno Kessler - Trento, Italy}
\affil[3]{Istituto Nazionale di Fisica Nucleare, Sezione di Firenze - Sesto Fiorentino, Italy}
\affil[4]{Fraunhofer USA Center Midwest - East Lansing, USA}
\affil[5]{Dipartimento di Fisica e Astronomia, University of Florence - Sesto Fiorentino, Italy}
\affil[6]{Laboratory of Nanooptics and Plasmonics, Moscow Institute of Physics and Technology - Dolgoprudny, Russian Federation}
\affil[7]{National Institute of Optics (INO), National Research Council (CNR) - Florence, Italy}

\begin{document}

\maketitle

\begin{abstract}
Phosphorus-doped diamond is relevant for applications in sensing, optoelectronics and quantum photonics, since the unique optical properties of color centers in diamond can be combined with the n-type conductivity attained by the inclusion of phosphorus. Here, we investigate the photoluminescence signal of the nitrogen-vacancy and silicon-vacancy color centers in phosphorus-doped diamond as a function of temperature starting from ambient conditions up to about 100$^\circ$ Celsius, focusing on the zero-phonon line (ZPL). We find that the wavelength and width of the ZPL of the two color centers exhibit a comparable dependence on temperature, despite the strong difference in the photoluminescence spectra. Moreover, the temperature sensitivity of the ZPL of the silicon-vacancy center is not significantly affected by phosphorus-doping, as we infer by comparison with silicon-vacancy centers in electronic-grade single-crystal diamond.
\end{abstract}

\section{Introduction}

Color centers in diamond are relevant for applications in sensing, optoelectronics and, more recently, also in quantum photonics~\cite{sipahigil_integrated_2016,schroder_quantum_2016,aharonovich_diamond_2018}. For instance, the spin-dependent fluorescence of the nitrogen-vacancy (NV) center is being exploited for sensing many physical parameters, including magnetic fields and temperature~\cite{acosta_temperature_2010,degen_quantum_2017}; electroluminescence devices based on p-i-n diodes have been developed down to single-photon emission ~\cite{lohrmann_diamond_2011,mizuochi_electrically_2012,tegetmeyer_electroluminescence_2016} and indistinguishable sources of single photons and coherent coupling of silicon-vacancy (SiV) color centers in diamond have already been demonstrated~\cite{sipahigil_indistinguishable_2014,evans_photon-mediated_2018}.

The temperature dependence of color centers in diamond is relevant for sensing, but also for light-emitting devices. It has recently been demonstrated that the performances of single-photon emission under electrical pumping largely improve as temperature increases~\cite{fedyanin_ultrabright_2016-1}.  Another important aspect is the creation of color centers in p- or n-type diamond, because they can be exploited in combination with the electrical conductivity of the host matrix. We have recently shown procedures for the creation of single SiV in phosphorus-doped (P-doped) diamond~\cite{flatae_silicon-vacancy_2020} and studied the temperature dependence of the SiV in intrinsic diamond at high temperatures~\cite{lagomarsino_robust_2015}. Next, the temperature sensitivity of the SiV in the biological range has attracted attention~\cite{nguyen_all-optical_2018,choi_ultrasensitive_2019}.

Here, we study the temperature sensitivity of the zero-phonon line (ZPL) of the negatively-charged NV and SiV in P-doped (n-type) diamond, showing that both centers can be used for all-optical temperature measurements at room and higher temperatures with comparable sensitivity. Moreover, we compare our findings with the temperature dependence of the SiV in optical-grade intrinsic diamond, demonstrating that the behavior of the SiV in n-type and intrinsic diamond as a function of temperature is similar.

\section{Results}

A P-doped 2-$\mu$m-thick single-crystal diamond film was grown on a high-pressure-high-temperature  diamond substrate by a homebuilt 2.45 GHz microwave plasma-enhanced chemical vapor deposition  reactor at a pressure of 160 Torr and 2 kW absorbed microwave power using phosphine (PH$_3$) as a dopant source. The plasma contained 0.09\% of CH$_4$, with a PH$_3$/CH$_4$ ratio of 4300 ppm. The intrinsic sample is a commercial optical-grade single-crystal diamond film from Element Six. Both samples have been implanted with Si ions at a fluence of 10$^{14}$ cm$^{-2}$ and thermally annealed to activate both centers~\cite{lagomarsino_center_2018}.

The optical experiments are performed with a home-built setup with a temperature-controlled sample holder and a 532 nm (690 nm) CW diode laser for the NV (SiV). The fluorescence signal is collected using a microscope objective (0.6 NA) and after long pass filters it is sent to the spectrometer.

\begin{figure}[h]
\includegraphics[width=\textwidth]{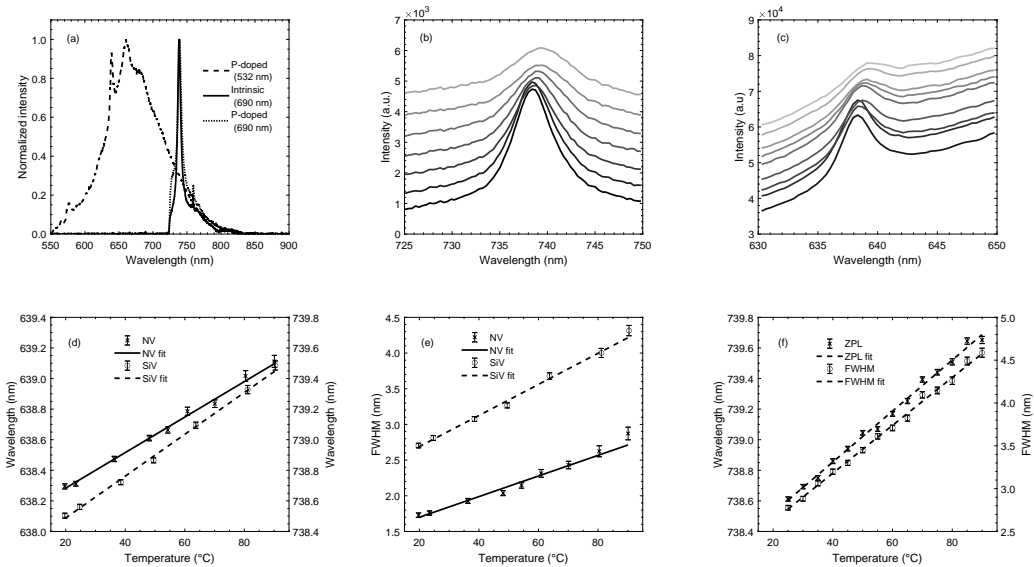}     
\caption{(a) Normalized PL spectra of the NV (dashed curve) and SiV (dotted curve) in P-doped and SiV (solid curve) in intrinsic diamond. PL spectra near the ZPL of the SiV (b) and NV (c) in P-doped diamond from room temperature to 90 $^\circ$C. Dependence of the ZPL center wavelength (d) and of the ZPL FWHM (e) with temperature for the SiV and NV in P-doped diamond and for the SiV in intrinsic diamond (f). \label{fig1}}
\end{figure}

Figure~\ref{fig1}a shows the normalized PL spectra of the NV (dashed curve) and SiV (dotted curve) in P-doped diamond as well as the spectrum of the SiV (solid curve) in intrinsic diamond  acquired at room temperature. The ZPL of both centers is clearly visible. When the temperature increases, the ZPL is redshifted and its linewidth increases (see Fig.~\ref{fig1}b and c for P-doped diamond). The amplitude of the ZPL decreases, but it remains visible up to higher temperatures. However, the ZPL of the NV disappears at temperatures above 200 $^\circ$C, whereas the ZPL of the SiV has been measured up to about 600 $^\circ$C~\cite{lagomarsino_robust_2015}.

Figure~\ref{fig1}d and e display the temperature dependence of the ZPL center wavelength and of the full-width at half-maximum (FWHM) for the NV and SiV in P-doped diamond. These values have been obtained from Fig.~\ref{fig1}b and c by fitting the ZPL with a Lorentzian profile. Next, the experimental data have been fitted with a weighted linear regression to obtain figure of merits for the temperature sensitivity. We remark that a linear fit would not be appropriate in a larger temperature range~\cite{lagomarsino_robust_2015}.

The temperature sensitivity for the NV and SiV is comparable and the fitting results read $\lambda_\mathrm{NV}=(638.052 \pm 0.017)$ nm + $T(11.6\pm 0.4)$ pm/$^\circ$C, $\lambda_\mathrm{SiV}=(738.215 \pm 0.017)$ nm + $T(13.7\pm 0.4)$ pm/$^\circ$C for the ZPL center wavelength and $\Delta\lambda_\mathrm{NV}=(1.41 \pm 0.04)$ nm + $T(14.5\pm 0.8)$ pm/$^\circ$C, $\Delta\lambda_\mathrm{SiV}=(2.25 \pm 0.04)$ nm + $T(21.8\pm 0.8)$ pm/$^\circ$C for the ZPL FWHM, where $T$ is the temperature in $^\circ$C. The SiV ZPL is slightly more sensitive than the NV ZPL and the FWHM of the ZPL is more sensitive than its center wavelength for both centers. Therefore, also the ZPL of the NV centers can be used for all-optical temperature sensing, such that the ODMR signal can be dedicated to magnetic field sensing.

Figure~\ref{fig1}f shows the temperature dependence of the SiV ZPL center wavelength and its FWHM  in intrinsic diamond. The linear fit yields $\lambda_\mathrm{SiV}=(738.187 \pm 0.013)$ nm + $T(16.7\pm 0.3)$ pm/$^\circ$C and  $\Delta\lambda_\mathrm{SiV}=(2.06 \pm 0.03)$ nm + $T(27.9\pm 0.5)$ pm/$^\circ$C. By comparison with the results of P-doped diamond, we can conclude that the presence of phosphorus does not significantly affect the temperature sensitivity of the SiV. A comparison with the NV in intrinsic diamond is not shown here, because the sample has a very low nitrogen content. Moreover, it is apparent that for sensing applications based on the ZPL of the NV a large presence of N in diamond is required, since most of the PL signal is found in the phonon sideband.

\section{Conclusions}

We reported PL studies of the NV and SiV in P-doped diamond, showing that the temperature sensitivity of the ZPL of the two centers is comparable, with a linear dependence in terms of wavelength and linewidth between ambient conditions and about 100 $^\circ$C. Moreover, our findings for the SiV are compatible with the behavior in intrinsic diamond. However, it is also evident that further studies on how doping affects the temperature response of color centers in diamond are required for applications in sensing and optoelectronics devices. 

\section*{Acknowledgments}

The authors acknowledge funding from the University of Siegen, the German Research Association (DFG) (INST 221/ 118-1 FUGG, 410405168) and the Russian Foundation for Basic Research (19-57-12008). G. Speranza and S. Piccolomo would like to thank the Provincia Autonoma di Trento for supporting their research in the framework of Q@TN initiative on Quantum Technologies. The authors also acknowledge INFN-CHNet, the network of laboratories of the INFN for cultural heritage, for support and precious contributions in terms of instrumentation and personnel. A.M. Flatae and M. Agio would like to thank P. Reuschel and N. Soltani for experimental support.


\begin{thebibliography}{10}

\bibitem{sipahigil_integrated_2016}
A.~Sipahigil, R.~E. Evans, D.~D. Sukachev, M.~J. Burek, J.~Borregaard, M.~K.
  Bhaskar, C.~T. Nguyen, J.~L. Pacheco, H.~A. Atikian, C.~Meuwly, R.~M.
  Camacho, F.~Jelezko, E.~Bielejec, H.~Park, M.~Lončar, and M.~D. Lukin, ``An
  integrated diamond nanophotonics platform for quantum-optical networks,''
  {\em Science}, vol.~354, no.~6314, p.~847, 2016.

\bibitem{schroder_quantum_2016}
T.~Schröder, S.~L. Mouradian, J.~Zheng, M.~E. Trusheim, M.~Walsh, E.~H. Chen,
  L.~Li, I.~Bayn, and D.~Englund, ``Quantum nanophotonics in diamond,'' {\em Journal of the Optical Society of America B},
  vol.~33, no.~4, pp.~B65--B83, 2016.

\bibitem{aharonovich_diamond_2018}
I.~Aharonovich and E.~Neu, ``Diamond {Nanophotonics},'' {\em Advanced Optical
  Materials}, vol.~2, pp.~911--928,  2018.

\bibitem{acosta_temperature_2010}
V.~M. Acosta, E.~Bauch, M.~P. Ledbetter, A.~Waxman, L.-S. Bouchard, and
  D.~Budker, ``Temperature {Dependence} of the {Nitrogen}-{Vacancy} {Magnetic}
  {Resonance} in {Diamond},'' {\em Physical Review Letters}, vol.~104,
  p.~070801,  2010.

\bibitem{degen_quantum_2017}
C.~L. Degen, F.~Reinhard, and P.~Cappellaro, ``Quantum sensing,'' {\em Rev.
  Mod. Phys.}, vol.~89, p.~035002,  2017.

\bibitem{lohrmann_diamond_2011}
A.~Lohrmann, S.~Pezzagna, I.~Dobrinets, P.~Spinicelli, V.~Jacques, J.-F. Roch,
  J.~Meijer, and A.~M. Zaitsev, ``Diamond based light-emitting diode for
  visible single-photon emission at room temperature,'' {\em Appl. Phys.
  Lett.}, vol.~99, no.~25, pp.~251106, 2011.

\bibitem{mizuochi_electrically_2012}
N.~Mizuochi, T.~Makino, H.~Kato, D.~Takeuchi, M.~Ogura, H.~Okushi, M.~Nothaft,
  P.~Neumann, A.~Gali, F.~Jelezko, J.~Wrachtrup, and S.~Yamasaki,
  ``Electrically driven single-photon source at room temperature in diamond,''
  {\em Nat. Photon.}, vol.~6, no.~5, pp.~299--303, 2012.

\bibitem{tegetmeyer_electroluminescence_2016}
B.~Tegetmeyer, C.~Schreyvogel, N.~Lang, W.~Müller-Sebert, D.~Brink, and C.~E.
  Nebel, ``Electroluminescence from silicon vacancy centers in diamond
  p–i–n diodes,'' {\em Diamond and Related Materials}, vol.~65, pp.~42--46,
   2016.

\bibitem{sipahigil_indistinguishable_2014}
A.~Sipahigil, K.~Jahnke, L.~Rogers, T.~Teraji, J.~Isoya, A.~Zibrov, F.~Jelezko,
  and M.~Lukin, ``Indistinguishable {Photons} from {Separated}
  {Silicon}-{Vacancy} {Centers} in {Diamond},'' {\em Physical Review Letters},
  vol.~113, p.~113602, 2014.

\bibitem{evans_photon-mediated_2018}
R.~E. Evans, M.~K. Bhaskar, D.~D. Sukachev, C.~T. Nguyen, A.~Sipahigil, M.~J.
  Burek, B.~Machielse, G.~H. Zhang, A.~S. Zibrov, E.~Bielejec, H.~Park,
  M.~Lončar, and M.~D. Lukin, ``Photon-mediated interactions between quantum
  emitters in a diamond nanocavity,'' {\em Science}, vol.~362, no.~6415,
  p.~662, 2018.

\bibitem{fedyanin_ultrabright_2016-1}
D.~Y. Fedyanin and M.~Agio, ``Ultrabright single-photon source on diamond with
  electrical pumping at room and high temperatures,'' {\em New Journal of
  Physics}, vol.~18, p.~073012,  2016.

\bibitem{flatae_silicon-vacancy_2020}
A.~M. Flatae, S.~Lagomarsino, F.~Sledz, N.~Soltani, S.~S. Nicley, K.~Haenen,
  R.~Rechenberg, M.~F. Becker, S.~Sciortino, N.~Gelli, L.~Giuntini,
  F.~Taccetti, and M.~Agio, ``Silicon-vacancy color centers in phosphorus-doped
  diamond,'' {\em Diamond and Related Materials}, vol.~105, p.~107797,
  2020.

\bibitem{lagomarsino_robust_2015}
S.~Lagomarsino, F.~Gorelli, M.~Santoro, N.~Fabbri, A.~Hajeb, S.~Sciortino,
  L.~Palla, C.~Czelusniak, M.~Massi, F.~Taccetti, L.~Giuntini, N.~Gelli, D.~Y.
  Fedyanin, F.~S. Cataliotti, C.~Toninelli, and M.~Agio, ``Robust luminescence
  of the silicon-vacancy center in diamond at high temperatures,'' {\em AIP
  Advances}, vol.~5, p.~127117, 2015.

\bibitem{nguyen_all-optical_2018}
C.~T. Nguyen, R.~E. Evans, A.~Sipahigil, M.~K. Bhaskar, D.~D. Sukachev, V.~N.
  Agafonov, V.~A. Davydov, L.~F. Kulikova, F.~Jelezko, and M.~D. Lukin,
  ``All-optical nanoscale thermometry with silicon-vacancy centers in
  diamond,'' {\em Applied Physics Letters}, vol.~112, p.~203102, 2018.

\bibitem{choi_ultrasensitive_2019}
S.~Choi, V.~N. Agafonov, V.~A. Davydov, and T.~Plakhotnik, ``Ultrasensitive
  {All}-{Optical} {Thermometry} {Using} {Nanodiamonds} with a {High}
  {Concentration} of {Silicon}-{Vacancy} {Centers} and {Multiparametric} {Data}
  {Analysis},'' {\em ACS Photonics}, vol.~6, pp.~1387--1392, 2019.

\bibitem{lagomarsino_center_2018}
S.~Lagomarsino, S.~Sciortino, N.~Gelli, A.~M. Flatae, F.~Gorelli, M.~Santoro,
  M.~Chiari, C.~Czelusniac, M.~Massi, F.~Taccetti, M.~Agio, and L.~Giuntini,
  ``The center for production of single-photon emitters at the
  electrostatic-deflector line of the {Tandem} accelerator of {LABEC}
  ({Florence}),'' {\em Nuclear Instruments and Methods in Physics Research B}, vol.~422, pp.~31--40,
  2018.

\end{thebibliography}

\end{document}